# Design of light concentrators for Cherenkov telescope observatories

François Hénault, Pierre-Olivier Petrucci, Laurent Jocou
Institut de Planétologie et d'Astrophysique de Grenoble
Université J. Fourier, Centre National de la Recherche Scientifique, 38041 Grenoble – France

Bruno Khélifi, Pascal Manigot, Stéphane Hormigos
Laboratoire Leprince-Ringuet, Ecole Polytechnique, 91128 Palaiseau – France

Jürgen Knödlseder, Jean-François Olive, Pierre Jean
Institut de Recherche en Astrophysique et Planétologie, 31028 Toulouse– France

Michael Punch
Université Paris 7 Denis Diderot, 75205 Paris – France

**ABSTRACT**

The Cherenkov Telescope Array (CTA) will be the largest cosmic gamma ray detector ever built in the world. It will be installed at two different sites in the North and South hemispheres and should be operational for about 30 years. In order to cover the desired energy range, the CTA is composed of typically 50-100 collecting telescopes of various sizes (from 6 to 24-m diameters). Most of them are equipped with a focal plane camera consisting of 1500 to 2000 Photomultipliers (PM) equipped with light concentrating optics, whose double function is to maximize the amount of Cherenkov light detected by the photo-sensors, and to block any stray light originating from the terrestrial environment. Two different optical solutions have been designed, respectively based on a Compound Parabolic Concentrator (CPC), and on a purely dioptric concentrating lens. In this communication are described the technical specifications, optical designs and performance of the different solutions envisioned for all these light concentrators. The current status of their prototyping activities is also given.

## 1 INTRODUCTION

Since more than thirty years, Cherenkov telescopes have been constructed all over the world. They aim at collecting very faint UV-peaked radiations generated at ground level by high energy Gamma-rays interacting with our atmosphere. Hence they can be used to make celestial maps of the most energetic events occurring in the Universe. A schematic drawing of a Cherenkov collecting telescope is depicted in Figure 1. It essentially consists of a single dish (parabolic or otherwise) of modest optical quality comparable to solar concentrators, which focuses the UV/blue radiation onto a focal plane equipped with secondary light concentrators in front of an array of photo-sensing devices. Although this paper is primarily centered on the Cherenkov Telescope Array (CTA) project [1], it must be emphasized that the optical principles described herein are applicable to any past, present or future Cherenkov telescope observatories.

CTA will be the largest cosmic gamma ray detector ever built in the world. It will replace existing observatories like H.E.S.S. [2], VERITAS [3] and MAGIC [4]. It will be built by an international Consortium including those of the authors, cooperating with many other Research Institutes worldwide. It should be installed at two different sites in the North and South hemispheres and should be operational for about 30 years. In order to cover the desired energy range from a few tens of GeVs to several hundreds of TeVs, the CTA is composed of dozens of collecting telescopes of three different sizes: Large-Size Telescopes (LST) of 24-m, Medium-Size Telescopes (MST) of 12-m as shown in Figure 2, and Small-Size Telescopes (SST) of 6-m diameters. Each of these telescopes is equipped with a focal plane camera of several square meters area, which typically consists of 1500 or 2000 Photomultipliers (PM). Each PM is equipped with its own Light Concentrator (LC), whose two main functions are firstly to maximize the amount of energy collected by the non-contiguous photosensitive areas of the PMs, and secondary to block any stray light originating from the terrestrial environment. Here, two different optical solutions have been designed for the LCs, respectively based on a Compound Parabolic Concentrator (CPC, also known as Winston cone), and on a purely dioptric Concentrating Lens (CL).

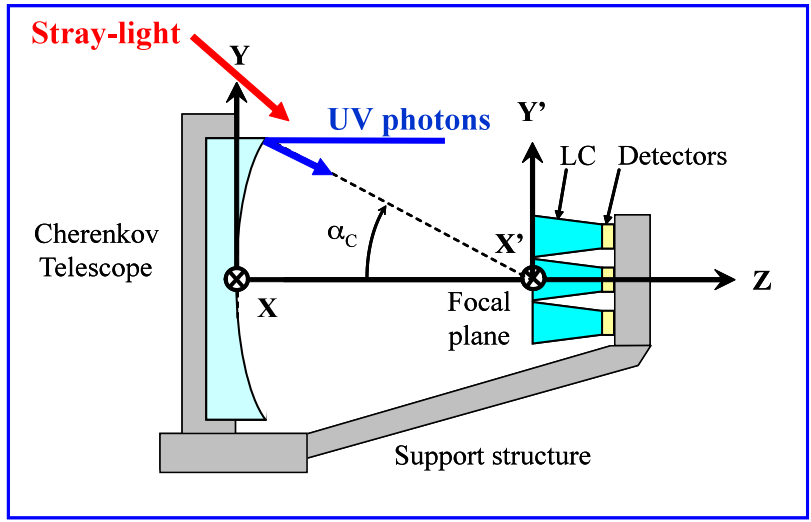

Figure 1: General layout of a Cherenkov light collecting telescope.

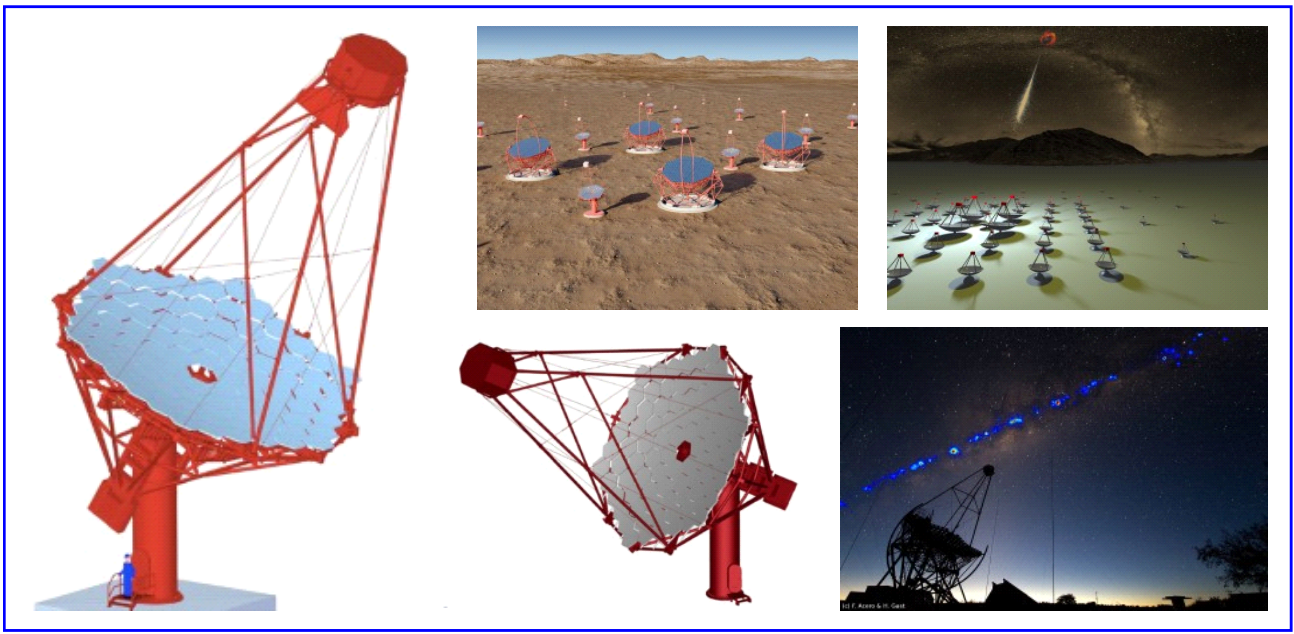

Figure 2: Tentative designs for the MST and some artist's view of the CTA observatory.

In this communication are described the technical specifications, optical designs and performances of the different solutions envisioned for the light concentrators that will equip the MST or LST focal planes (section 2). A short tentative trade-off between lenses and cones is discussed in section 3. The current status of prototyping activities is given in section 4 before summarizing the current status of the project in section 5.

## 2   OPTICAL DESIGN OF LIGHT CONCENTRATORS

### 2.1   Technical requirements

The major technical requirements for the MST light concentrators are given in Table 1. Two of them are essential:

- A spectral range of [300-600 nm], adapted to collecting most of the photons comprising the Cherenkov radiation where most of the energy lies in the UV domain.

- A cut-off angle $\alpha_C$ (see Figure 1) matched to the aperture of the observing telescope, in order to collect most of the useful photons without adding stray-light originating from night-sky background and terrestrial parasitic lights. The values of 26-28 deg. used here are derived from an analytic model optimizing the Signal-to-Noise Ratio (SNR) of the whole system[1]. More details about this SNR model are provided in the Appendix.

Other specifications such as the slopes of rejection curves at cut-off angle, or crosstalk between adjacent LCs must not be neglected. In addition, strong constraints may result from the curvature radius of the photo-sensitive area that is here 20 mm for the PMs selected for CTA. The whole set of requirements led us to design two different types of optical solutions, either reflective (§ 2.2) or dioptric (§ 2.3).

Table 1: Major technical requirements for Light Concentrators (LC).

| REQUIREMENTS | VALUES |
| --- | --- |
| Spectral range | From 300 to 600 nm |
| Optical transmission | ≥ 75 % on full spectral range (goal 80%) |
| Cut-off angle $\alpha_C$ (see Figure 1) | 26 deg. ≤ $\alpha_C$ ≤ 28 deg. |
| Rejection rate outside of cut-off angle $\alpha_C$ | ≤ 5 % on full spectral range |
| Crosstalk between two adjacent LCs | ≤ 3 % on full spectral range |
| LC entrance aperture | Hexagonal, ∅ = 50 mm flat to flat |
| Curvature radius of photo-detector (PM) | 20 mm |
| Diameter of photo-sensitive area (PM) | 34 mm |

---

[1] In opposition with astronomy imaging telescopes where long integration times are desirable in order to improve their SNR, the best performance of Cherenkov telescopes are achieved when integration time is matched to the duration of the Cherenkov pulses, i.e. a few tens of nanoseconds. Hence their SNR models do not follow the same logic.

## 2.2 The reflective solution: Winston cones

This all-reflective design is based on the utilization of CPCs (i.e. Winston cones) that was implemented on most existing Cherenkov observatories. They consist of axis-symmetric parabolic reflective sections, where the axis of the mother parabola is not aligned onto the centre of the Field of View (FoV) but rather to its edge: this is known as the "Edge-ray principle" explained in the reference textbook [5]. It follows that those concentrators actually are non-imaging optics. Furthermore, they obey a law stating that $\sin \alpha_C = y'' / y'$, where y' and y" respectively are the input and output diameters of the CPC. Then the parameter y" standing for the spot size on the PM photo-cathode should be around 24.5 mm, corresponding to an effective concentration factor $C = [y'/y'']^2 \approx 4.2$.

All computations were carried out using the Zemax™ commercial optical software [6], working in its non-sequential mode. Pictures of the ray-tracing and 3D view of the concentrator are reproduced in Figure 3. Although the theoretical length of the CPC is equal to $L = [y'+y''] / \tan \alpha_C \approx 66$ mm, it has been reduced to L = 54 mm in order to save volume and mass: this is actually a "truncated Winston cone" such as described in the Appendix I of Ref. [5]. Finally, a safety margin of at least 1 mm is imposed between the exit side of the CPC and the entrance face of the PM in order to prevent electro-static discharges.

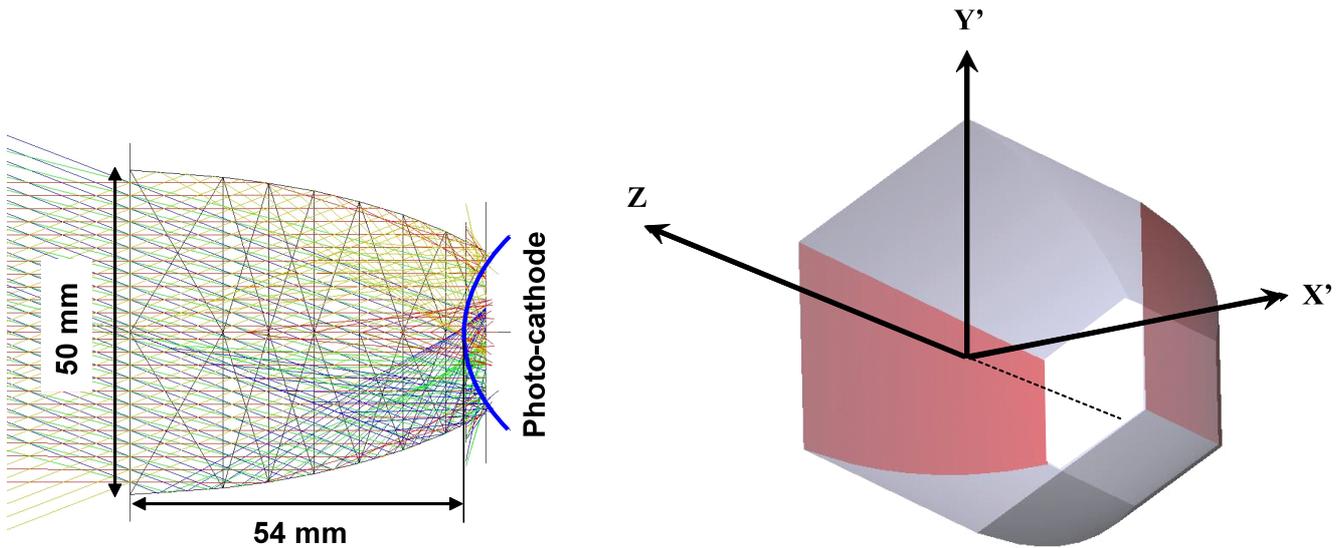

Figure 3: Ray-tracing and 3D view of the truncated Winston cone.

## 2.3 The dioptric solution: Concentrating lenses

In this section are described the methodology that was followed for defining a dioptric solution (§ 2.3.1) and the resulting optical designs (§ 2.3.2).

### 2.3.1 Methodology

By "dioptric solution" we mean that the LCs should only be constituted of glass refractive materials. It was first intended to use Fresnel lenses, but it rapidly turned out that the high entrance aperture angle of the telescope would impose severe light losses due to shadowing between adjacent slices. The second idea was to design a classical field lens conjugating the telescope pupil to the entrance side of the PM. All computations were carried out using the Zemax software in a conventional sequential mode, and trying to optimize image quality over the full FoV. But here again, due to the large dimensions of the input FoV and LC diameters, the imaging quality was so poor that the concentrating requirements of Table 1 could hardly be met.

Our dioptric solution finally consists in neglecting the imaging quality criterion at the FoV centre, which is easily feasible using most of standard optical software. To illustrate the point, Figure 4 depicts two footprint diagrams on the PM photo-cathode sensitive area computed via Zemax. They exhibit fully defocused spots at the FoV centre (red dots) while the spot sizes can be minimized at FoV edge (green and blue dots). These kinds of non-imaging lenses are not designed to form clear images of the full telescope pupil, but only of its rim. This is just another application of the Winston's edge-ray principle [5] applied to dioptric systems. Moreover, it also shows that common optical software may be useful for pre-designing non-imaging optics systems, and providing first-guess solutions in a quick way. Obviously, such preliminary solutions could later be optimized using more refined techniques described in Ref. [5]. The optical designs resulting from this first step are described in the next sub-section.

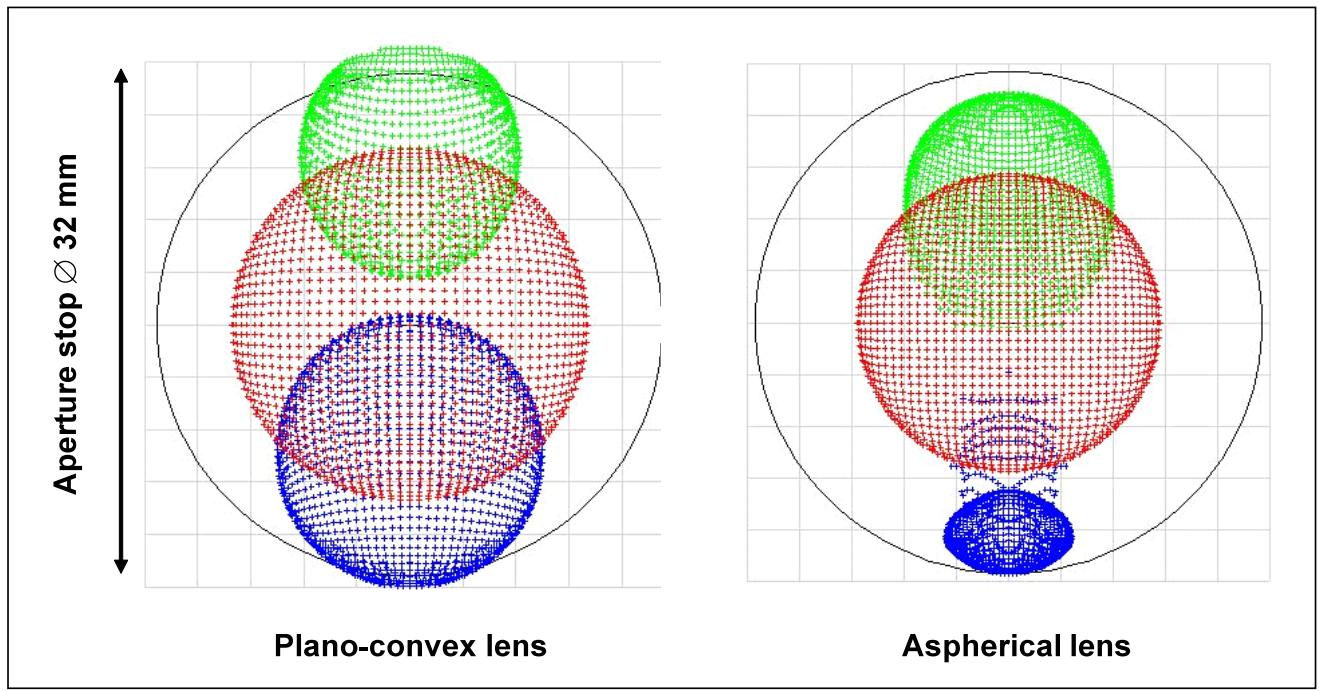

Figure 4: Footprint diagrams of non-imaging lenses at the PM entrance window. Red dots: footprint at FoV center. Green and blue dots: footprints at FoV edges.

*2.3.2 Optical designs*

We designed two different non-imaging lenses complying with the requirements of Table 1. These optical solutions are illustrated by ray-tracing and 3D views in Figure 5 and consist in:

- A plano-convex lens of hexagonal section, whose entrance face (on telescope side) is curved and exit face (on PM side) is flat.
- An aspherical lens that was optimized starting from the previous design: here the exit face has been slightly aspherized as can be seen on the bottom left panel of Figure 5.

For these lenses we selected the FK5 Schott glass material (but other equivalent glasses are available from different manufacturers, e.g. Ohara or CDGM). This glass is cheap, has a good transmission in the UV range and presents the advantage of being moldable, therefore allowing significant cost reductions in the mass production phase (other choices such as Fused Silica or CaF2 would have led to unaffordable costs). Both lens faces are anti-reflective coated with a standard $\lambda/4$ MgF2 layer in order to improve the transmission. Finally, the concentrating lenses are mounted into a baffle ending on an aperture stop on the PM side (see sub-section 4.2), playing exactly the same role as the exit aperture of the Winston cone. The diameter of the spot on the entrance side of the PM is around 32 mm, which corresponds to a concentration ratio $C \approx 2.5$ (here again a safety margin of 1 mm separates the entrance face of the PM and the stop). The basic performance of both the dioptric and reflective systems will be discussed in the next section.

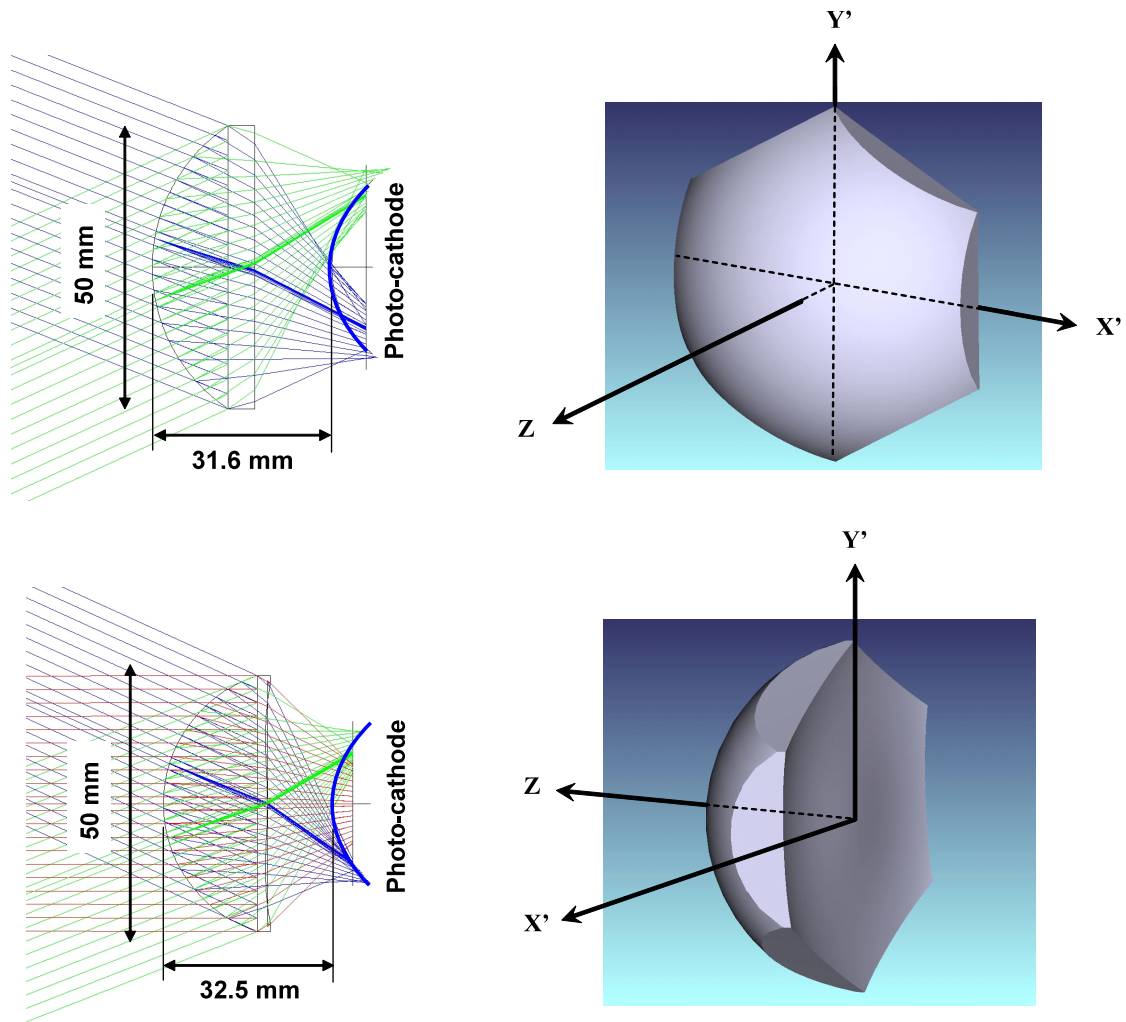

Figure 5: Ray-tracing and 3D views of the plano-convex (top) aspherical (bottom) non-imaging lenses.

## 2.4 Expected performance

We consider that the most important performance of the LCs is characterized by their absolute transmission (e.g. at FoV center) and the slopes of their rejection curves at cut-off angle $\alpha_C$. With the help of Zemax software, the rejection curves of the hexagonal CPC, plano-convex lens and aspherical lenses were computed and are shown in Figure 6. It can be seen that the sharpest slopes are achieved by the non-imaging lenses, and particularly by their aspherical version. This is one of the reason why, for achieving equivalent SNRs, the lens cut-off angle $\alpha_C$ (here equal to 26 degs.) could be reduced with respect to that of the cone (28 degs.) Moreover, the presence of anti-reflective coatings on the lens faces clearly improves their transmission, whereas the reflective coating of the cone mainly consists of two thin layers deposited on its plastic shell, and seen under high angles of incidence. However, a more complete trade-off needs to be done at this stage, and further advantages and drawbacks of both types of solutions will be discussed in section 3.

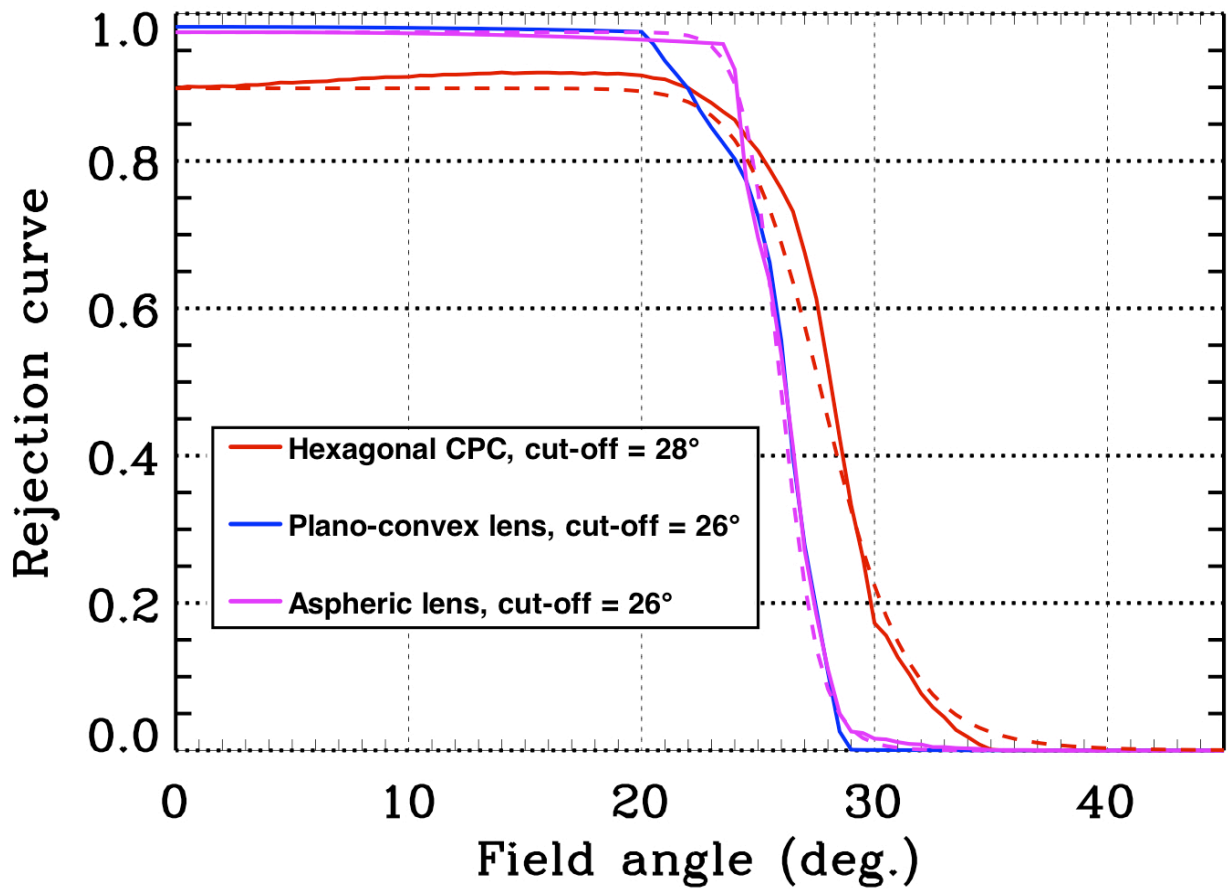

Figure 6: Typical rejection curves achievable by the hexagonal CPC and non-imaging lenses. Signification of the red and magenta dashed lines is explained into the Appendix.

## 3   CONES vs. LENS TRADE-OFF

The main elements of the trade-off are summarized in Table 2. Some criteria featuring in this Table are self-explanatory (e.g. weight and cost per individual unit) or have already been discussed in section 2 (transmission, slope of rejection curves…). A few other deserve further comments.

Table 2: Cone vs. non-imaging lens trade-off summary.

| Criterions | HEXAGONAL CONE | CONCENTRATING LENS |
|---|---|---|
| Typical cost (per unit) | 15 Euros | 20-30 Euros |
| Weight (per unit) | 15 grams | 40-50 grams |
| Size of entrance area | Not critical | Should be ≤ 50 mm |
| Need for protective window | Yes | No (lens acting as window) |
| Cut-off angle for straylight rejection | 28 deg. | 26 deg. |
| Maximal slope of rejection curve | 12% / deg. | 3% / deg. |
| Typical absolute transmission | 0.8 | 0.96 (with AR coatings) |
| Alignment tolerance (wrt PM) | Moderate | Moderate |
| Spot size on cathode | 24 mm | 32 mm |
| Dead spaces in focal plane | ≤ 1 mm | ≤ 1 mm |
| Encumbrace (along optical axis) | 54 mm | ≈ 32 mm |

- Size of entrance area: Our study revealed that the dioptric solution is more suited to small LC entrance diameters, while the reflective solutions are not affected by such an inconvenience. This is due to the fact that, even for non-imaging lenses, high-aperture and FoV aberrations (e.g. spherical aberration or coma) can notably degrade image quality at FoV edges.

- Need for protective window: This is perhaps one of the most critical items in this trade-off, because previous experience [2-4] demonstrated that dusty or sandy environments could be harmful for Winston cones and PM entrance areas. Hence several technical solutions could be envisaged:

  a) To protect the whole camera assembly with a large Plexiglas window covering the full telescope focal plane. This solution presents the advantage that it could be incorporated into the camera cooling system, but has the drawback that the window should be thick enough to limit mechanical flexures, therefore strongly decreasing the number of collected UV photons.

  b) For the reflective solution, to add individual, very thin Plexiglas windows at the entrance surface of Winston cone. This is an elegant solution, but makes the LC opto-mechanical design more difficult.

  c) Finally, to take benefit of the dioptric design, considering lenses themselves as natural protective windows.

- Alignment tolerance: In both cases, this was not found to be very critical, being of the order of a few tenths of millimeters in translation and one degree in rotation.

- Dead spaces between LCs: These result from lost areas between adjacent LCs, originating either from cone shell thickness or lens external supports. Mechanical studies showed that these losses are equivalent.

- Encumbrance along Z-axis: It should be noted that the back-focus length of lenses (around 32 mm) is significantly lower than for the CPC (54 mm), which gives them an advantage in terms of allocated volume.

In the end, it seems clear that the conclusion of this trade-off is far from obvious. This is the reason why IPAG, the "Laboratoire Leprince-Ringuet" (LLR) and the "Institut de Recherche en Astrophysique et Planétologie" (IRAP) undertook the manufacturing of prototype light concentrators and developed together a dedicated test bench located in Toulouse (France), having the capacity to evaluate both technical solutions on the basis of their rejection curves.

## 4 PROTYPING ACTIVITIES

### 4.1 Winston cones

The practical realization of prototype Winston cones was assigned to the Savimex company located in Grasse (France), based on their previous experience on the H.E.S.S. observatory. Most of the achievements are illustrated in Figure 8. On the top left panel is shown a CAD view of the LC. The right side panel shows a picture of the assembled prototype that consists of a parabolic plastic shell coated with high-reflective layers (Al+MgF2). The bottom picture illustrates the mounting scheme of this LC that is constructed by assembling three couples of molded parabolic segments in order to reduce non-homogeneities during the coating process. Finally, the bottom side of this picture shows a hexagonal Plexiglas protective window that should be clipped on the entrance of the LC.

### 4.2 Concentrating lenses

Both types of non-imaging lens prototypes (plano-convex and aspheric) have been ordered and delivered from two different manufacturers. From left to right, Figure 7 depicts the theoretical scheme of the system (including baffle, aperture stop, and the curved PM photo-cathode – not to scale), a CAD view of the non-imaging lens inserted into its baffle, and one picture of the manufactured and assembled sub-system.

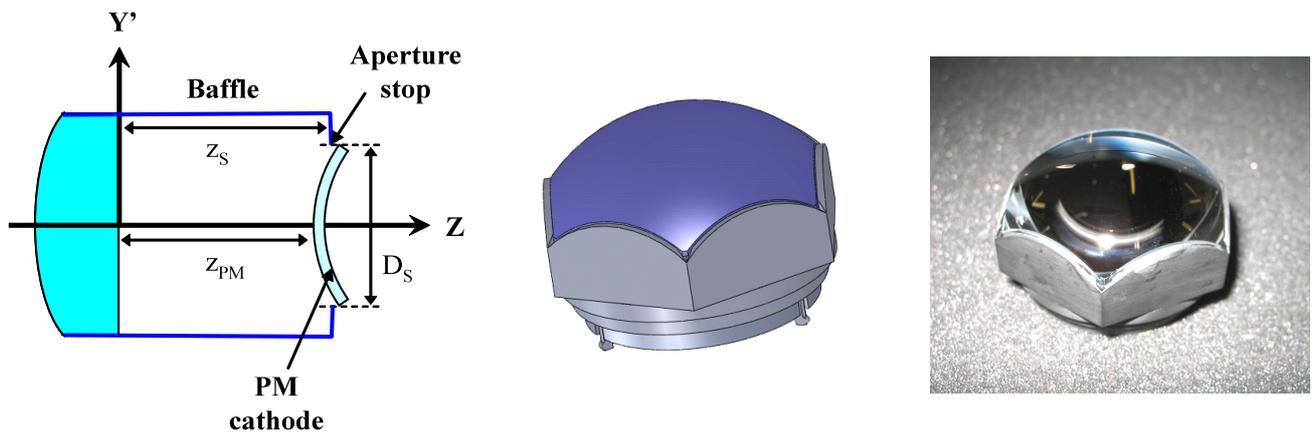

Figure 7: CAD view and pictures of the prototyped hexagonal non imaging lens and its baffle.

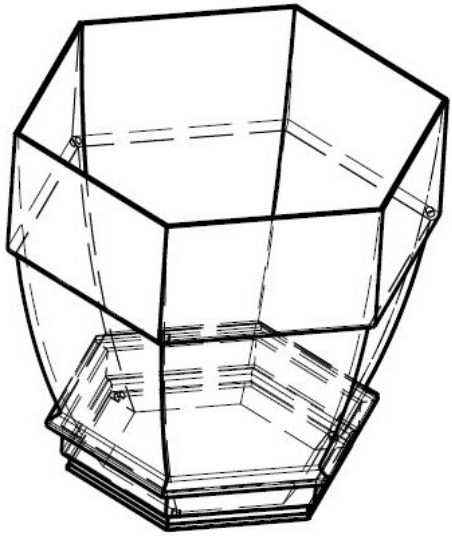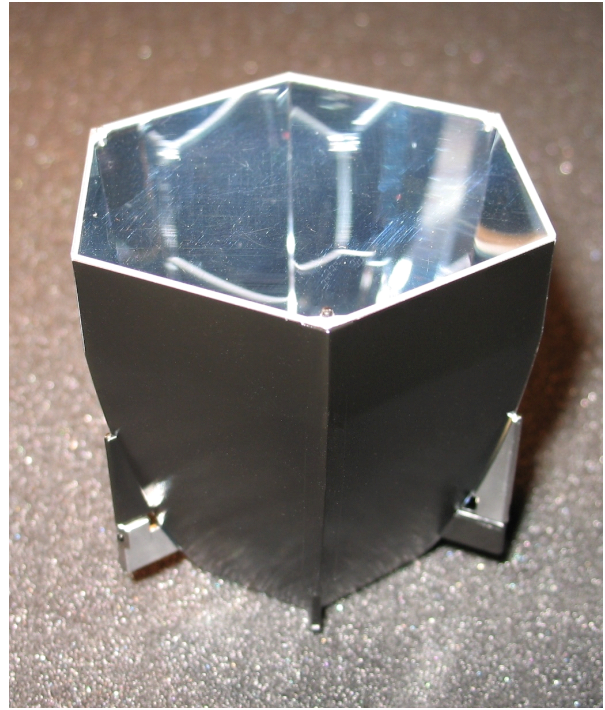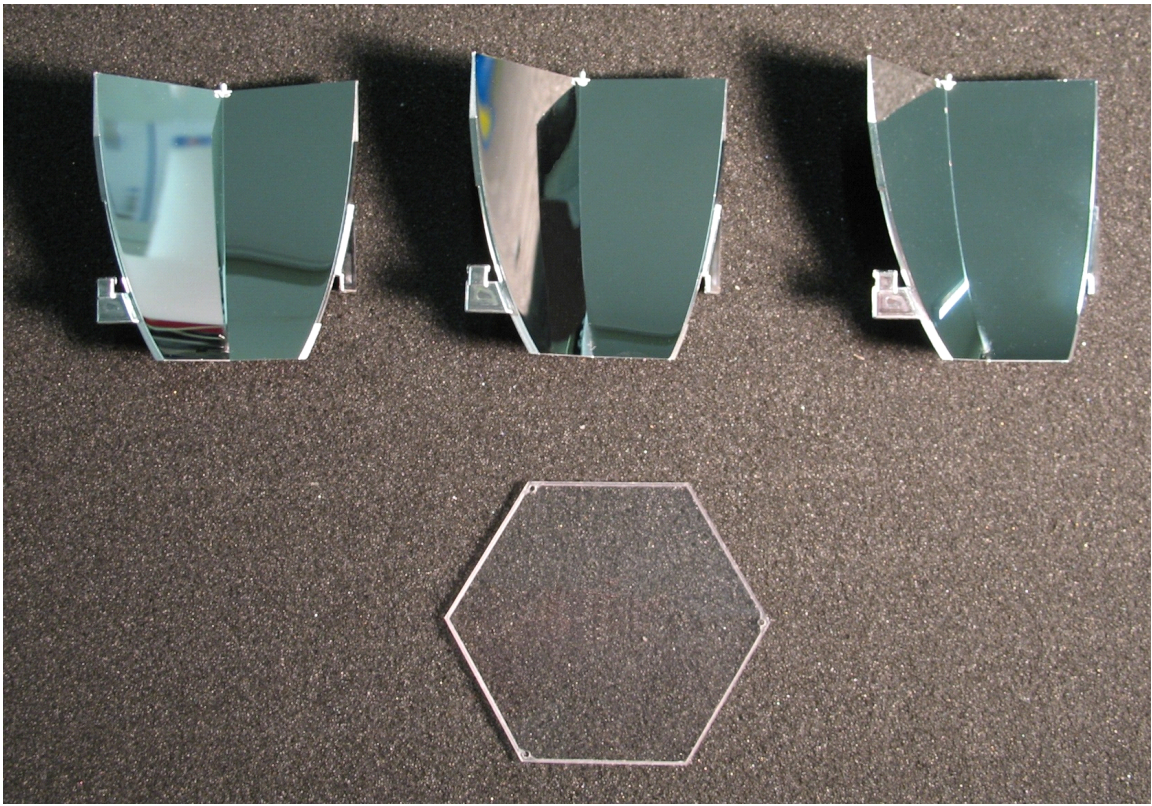

Figure 8: CAD view and pictures of the prototyped Winston cones and protective window.

## 5 PROJECT STATUS AND CONCLUSION

In this paper were summarized our studies about light concentrators for the Cherenkov Telescope Array (CTA), that could be extended to any Cherenkov telescope observatory. Here we identified two optical solutions based on Winston cones and on non-imaging lenses. Today their prototypes have been manufactured and delivered to IPAG for visual inspection, then sent to IRAP laboratory for preliminary testing on the optical bench. Figure 9 shows a first measurement result of a prototyped CPC rejection curve performed at IRAP. This is rough data, being not calibrated and to be subject to further processing. However, it already reveals some similarities with the theoretical curve of Figure 6 (e.g. a transmission "bump" near FoV edge).

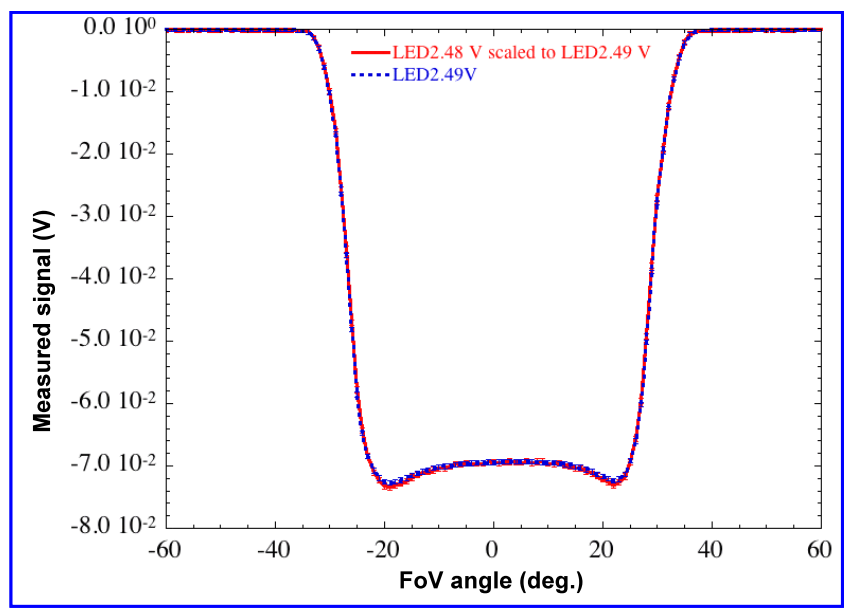

Figure 9: An example of first CPC rejection curve measured at IRAP.

An extensive test campaign should begin at IRAP in September 2013, allowing us to evaluate the major performance criteria of different LC types, and to answer to some open questions discussed in section 3. A final conclusion is expected at the end of year 2013. In parallel, other studies will be carried out for adapting the reflective or dioptric designs to the cases of LST (Large-Size Telescopes) and SST (Small-Size Telescopes) for the CTA observatory.

**Acknowledgements:** This work has been supported by grants from Labex OSUG@2020 (Investissements d'avenir - ANR10 LABX56) and fundings from the European Union's Seventh Framework Programme (FP7/2007-2013) under grant agreement n° 262053

# APPENDIX. Optimization of LC cut-off angle

The cut-off angle $\alpha_C$ (see Figure 1) of the light concentrator should be optimized in order to collect onto the PM photocathode most of the useful Cherenkov photons collected by the telescope without adding stray-light originating from night-sky background (NSB) and terrestrial parasitic lights. For that purpose, at a given pixel position on the camera, we defined the following optimization function:

$$\text{SNR}(\alpha_c) = \frac{M_{Ch}(\alpha_c)}{\sqrt{M_{NSB}(\alpha_c) + aA_{NSB}(\alpha_c)}} \quad (1)$$

where $M_{Ch}$ is the total number of Cherenkov photons directly reflected by the collecting telescope, $M_{NSB}$ the number of NSB photons coming from the telescope mirror, $A_{NSB}$ the number of NSB photons impinging at ground level, and $a$ the ground albedo coefficient. The estimates of $M_{Ch}$, $M_{NSB}$ and $A_{NSB}$ require to take into account the complete environment of the light concentrator like the telescope characteristics (mirror diameter and F/D number, camera FoV and pixel size, mirror reflectance, etc...), the PM characteristics (quantum efficiency, entrance window size and shape) as well as the Cherenkov and NSB spectral properties (spectral shape, flux, mean Cherenkov shower shape and duration, etc...). A more detailed description of the computation of this optimization function should be published elsewhere.

The results shown here assume a MST telescope and the Hamamatsu PM R11920-100-02. We also assume a Cherenkov shower produced by a 20 GeV photon, with a duration of 10 ns and a mean size of the shower on the camera focal plane of about 10 pixels. Finally, we choose a camera pixel at the edge of the camera FoV (i.e. 4 deg. off axis in the case of MST). To speed-up the computation process, we approximate the LC angular rejection curves with super-Lorenz profiles (see the dashed lines in Figure 6). These profiles mimic relatively well the ones we obtain with the Zemax simulations. Then, for a given cut-off angle $\alpha_C$ between 0 and 45 degrees, we compute the optimization function SNR($\alpha_C$) taking into account all the parameters indicated above. The results are shown in Figure 10. On the left side of this Figure, we have plotted the SNR curves for the cases of an hexagonal Winston cone and an hexagonal aspherical lens, assuming an albedo a = 0.2. In both cases, the curves rapidly increase, reaching a maximum $\alpha_{Cmax}$ and then slightly decreasing. The maximum is reached at a slightly lower cut-off angle for lenses (~ 26 degs.) compared to CPC (~ 28 degs.), the main reason being the sharper profile of the lens rejection curve. The lens SNR curve is also always higher than the CPC one due to the better transmission of the lenses compared to reflective CPC.

On the right of Figure 10, we have plotted the values of $\alpha_{Cmax}$ for different values of albedos. The higher the albedo, the larger the number of photons coming from the albedo regions (the $A_{NSB}$ parameter in Eq. 1) and consequently the smaller $\alpha_{Cmax}$. For an albedo of 1, $\alpha_{Cmax}$ reaches 24 degs.

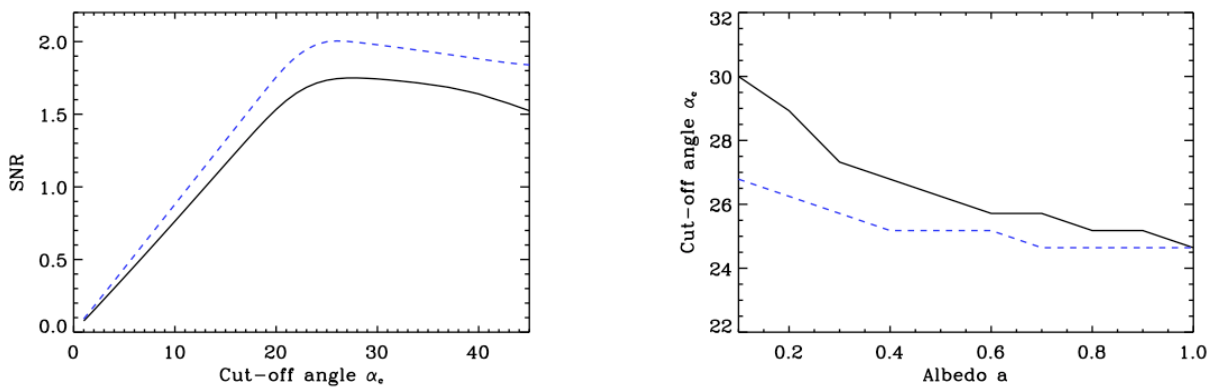

Figure 10: Left: the optimization functions SNR($\alpha_C$) for an aspherical lens (blue dashed line) and a CPC (black solid line) for an albedo of 0.2. Right: the angle $\alpha_C$ at which the SNR curve is maximal, in function of the albedo (blue dashed line for an aspherical lens and black solid line for a CPC).